\documentclass[paper=A4,pagesize,DIV=13,12pt]{scrartcl}

\usepackage[english]{babel}
\usepackage{csquotes}
\usepackage{eurosym}
\usepackage{graphicx}
\usepackage[affil-sl]{authblk}
\usepackage{amsmath}
\usepackage{amssymb}
\usepackage[tikz]{bclogo}
\usepackage{etoolbox}%
\usepackage{blindtext}%
\usepackage[skins,breakable,xparse]{tcolorbox}%
\usepackage{url}
\usepackage{blindtext}
\usepackage{lmodern}
\usepackage[T1]{fontenc}
\usepackage{microtype}
\usepackage{xcolor}
\usepackage{framed,color}
\usepackage{calc}
\usepackage{hyperref}
\usepackage{siunitx}
\usepackage{todonotes}
\usepackage[natbib=true,
    style=vancouver,
    sorting=none,
    url=false,
    maxbibnames=6,
    minbibnames=6]{biblatex}
\usepackage{tikz}
\usetikzlibrary{shapes,arrows,matrix,positioning,fit,backgrounds}
\graphicspath{{figures/}}
\usepackage{caption}
\usepackage{subcaption}
\usepackage{booktabs}
\usepackage{makecell}
\usepackage{fontawesome5}
\usepackage{orcidlink}
\usepackage{widetext}
\usepackage{cleveref}
\usepackage{siunitx}
\usepackage{eurosym}
\usepackage[version=4]{mhchem}
\usepackage[%
  automark,
  headsepline,                
  ]{scrlayer-scrpage}
  
\newcommand{\shorttitle}{Dummy}

\lohead{\sffamily Section~\headmark}    
\rohead{\pagemark}                        
\lofoot{\sffamily \shorttitle}
\rofoot{} 

\automark[subsection]{section}
\tcbset{colback=black!5!white,colframe=black}

\setkomafont{title}{\sffamily\bfseries}
\setkomafont{subtitle}{\normalfont\normalsize}
\addtokomafont{author}{\raggedright\setlength{\tabcolsep}{0pt}\normalsize}
\addtokomafont{date}{\raggedleft\normalsize}
\setkomafont{caption}{\slshape}
\setkomafont{captionlabel}{\normalfont\bfseries}

\usepackage{diagbox}
\usepackage{nicefrac}
\DeclareSIUnit{\EUR}{\text{€}}

\begin{document}


\title{Economical and ecological impact of sector coupling applied to computing clusters}
\renewcommand{\shorttitle}{}
\subtitle{}
\author[1]{Philip Bechtle}
\author[1]{Oliver Freyermuth}
\author[1,2]{Maximilian Peter Geffers}
\author[3]{Manuel Giffels}
\author[1]{Michael Hübner}
\author[1,*]{Florian Kirfel}
\author[4]{Jochen Kreutz}
\author[4,5]{Stefan Krieg}
\author[1]{Sebastian Matberg}
\author[3]{Matthias Schnepf}

\affil[1]{Physikalisches Institut, University of Bonn, Bonn, Germany}
\affil[2]{Universität Heidelberg, Heidelberg, Germany}
\affil[3]{Karlsruhe Institute of Technology, Karlsruhe, Germany}
\affil[4]{Jülich Supercomputing Centre, Forschungszentrum Jülich, Jülich, Germany}
\affil[5]{Helmholtz‑Institut für Strahlen- und Kernphysik, University of Bonn, Bonn, Germany}
\affil[*]{Contact: \texttt{fkirfel@uni-bonn.de}}
\date{\today{}}

\maketitle

\begin{abstract}
      \noindent\textbf{\sffamily Abstract:}
      The rising share of abundant renewable energy inevitably increases volatility in the electricity production. The concept of sector coupling means that the volatility of electricity production to a large degree can be absorbed by dispatching electricity consumption whenever excess renewable energy is available. A system that is dynamically operated based on this principle can lower its total environmental impact. In addition, operational costs might be reducible as electricity prizes strongly depend on the residual load of the energy system. High-performance computing clusters in the field of science represent an ideal testing ground for such dynamic operation. Short-term delays in computing results due to electricity production being associated with high costs or carbon emissions are often negligible, provided that an overall computing target remains constant over long time periods.
      This study simulates the simplified operation of computing clusters using publicly available data on electricity production in Germany. The optimal utilisation along with associated carbon emission and cost reductions are determined separately. Hardware acquisition costs and embedded emissions are taken into account. The stability of a fixed computing target given the determined utilisation optima is evaluated in two validation periods. Additional simulations with modified parameters are carried out to estimate potential conditions under which dynamic operation of a computing cluster would continue to enable savings in the future.      
\end{abstract}

\newpage

\tableofcontents

\pagestyle{scrheadings}


\section{Introduction}
\label{sec:intro}
The global warming induced by carbon emissions, first predicted in 1896 \cite{Arrhenius01041896}, is expected to cause a global increase in average temperatures of more than \qty{1.5}{\celsius} by 2030 compared to pre-industrial levels \cite{ctx19612175960006467}. The consequences of this trend are already evident across Europe in the form of more frequent and more intense extreme weather events. In Germany alone, the annual costs of recorded extreme-weather-related damages exceeds six billion euros during the period from \numrange{2015}{2024} \cite{stover2025kosten}.

To at least mitigate these effects in the future, Germany, originally along with 195 other nations, committed itself to the Paris climate agreement with the goal of limiting global warming to well below \qty{2}{\celsius} above pre-industrial levels \cite{paris}. This requires a drastic reduction in greenhouse gas emissions, primarily achievable through lowering the carbon intensity of the German electricity mix, which amounted to an average of \qty{344}{kg\ce{CO2}\per MWh} in 2025 \cite{Umweltbundesamt}, by switching to renewable energy sources. In 2025, the share of renewable energies in Germany electricity consumption was already at \qty{55.1}{\percent} \cite{BMWK}. However, its share in the final total energy consumption only amounted to \qty{23.8}{\percent} \cite{BMWK} and was thus significantly lower. This discrepancy can be attributed to sectors such as transport and heating, both of which still rely heavily on fossil fuels. 

Naturally, attempting to electrify these areas as well requires more energy to be produced by renewable energy sources, i.e.\ volatile wind and solar power will become the dominant power sources. Ultimately this increases fluctuations in the energy grid. To control such fluctuations, a strong sector coupling \cite{brown2018synergies} is required, capable of leveraging the flexibility of various sectors. In contrast to large industrial plants, which, to be profitable, must often operate continuously, the focus for sector coupling lies on systems that can be dynamically regulated. An almost perfect use case for this desired dynamic regularisation are computing clusters, as their workload can be flexibly controlled by shifting computing tasks over time. As of 2025 data centres already account for \qty{4}{\percent} of Germany's gross electricity consumption \cite{BMWK_DataCentreLandscape_2025} and, given the recent developments in the field of artificial intelligence, will account for an ever-increasing share of consumed power. Forecasts assume that, by 2030, computing might be responsible for up to \qty{8}{\percent} of the total power consumption \cite{cao2022systematicsurveycarbonneutral}. 

Shifting computing tasks to periods with an oversupply of renewable energies can help at stabilising the energy grid. At the same time, procuring electricity from the spot market offers significant savings in terms of carbon emissions, while also enabling economically optimised operation of computing clusters. As an example, negative wholesale energy prices on the German electricity spot market were reported for a total of 573 hours in 2025 \cite{SMARD_RecordHighSolar_2026}, primarily at weekends or on public holidays during which renewable energy sources led to an oversupply. Targeted ramp-up of a computing cluster at such periods therefore offers potential cost savings.

This dynamic operation of computing clusters inevitably leads to delays in computing results on short time scales. However, this is often negligible, especially in the field of science, making computing clusters in research institutions and universities the ideal place to test such an application. The following studies are therefore carried out by relying on different computing cluster operated by research facilities. The main focus is placed on the BAF high throughput computing cluster \cite{freyermuth2021operating}, operated at the Physics Institute of the University of Bonn. This cluster is primarily used for computing tasks in the field of high-energy physics, which, with more than \qty{20}{\percent} \cite{Suarez_2025}, has the greatest demand for computing power in the German scientific sector and thus also represents the largest lever for dynamic control. 

Initially, in \cref{sec:background} various options for reducing greenhouse gas emissions in the area of computing clusters are discussed. Particular emphasis is placed on the feasibility of the respective countermeasures on a short time scale. Building on this, \cref{sec:methodology} introduces the simulation of a dynamic cluster operation for data centres by relying on public data on Germany's electricity generation in recent years. Finally, \cref{sec:result} presents results for concrete case studies based on this simulation.



\section{Background}
\label{sec:background}

In general, computing clusters account for a significant share of total emissions in modern research. In particular, large high-energy physics experiments almost all rely on computing clusters. A study evaluating the professional carbon footprint of doctoral researchers at one of the four major LHC experiments shows that computing with \qty{1.91}{t\ce{CO2}} accounts for \qty{9.3}{\percent} of the experiment's total emissions \cite{lang2025know}. 

Fortunately, computing clusters offer many possibilities to reduce their carbon footprint either directly or indirectly. The first key decision for any newly built computing cluster involves selecting an adequate location. Besides the clearly important access to power supply, the local climate also plays a significant role. Depending on a cluster's architecture, current outside temperatures in Germany allow operation without active cooling \cite{suarez2024energyawareoperationhpcsystems}. This leads to a significant reduction of carbon emissions associated with the cluster operation. The heat, inevitably generated during operation, can be fed into the local heat networks of neighbouring towns. However, the temperature gradient between existing heat network's operating temperatures and the waste heat generated often poses a problem \cite{suarez2024energyawareoperationhpcsystems}. 

Regarding a computing cluster's installed hardware, using more modern components usually goes hand in hand with lower carbon emissions. This is either achieved through the components being more energy efficient or simply due to their increased performance, which shortens the overall runtime of computing tasks. Obviously, the embedded emissions associated with producing and transporting hardware have to be factored in. 

Ultimately, training end users also offers a great indirect potential for savings. Every end user initially must learn to correctly estimate the cluster resources required for their own computing tasks, e.g.\ the number of CPUs needed to efficiently complete a certain task. Furthermore, the correct use of an appropriate programming language is crucial to make optimal use of the available hardware. In large scientific collaborations in particular, further savings can be achieved by adopting the FAIR data principle \cite{wilkinson2016fair}. This principle intends to ensure that data and software are easily findable, accessible, interoperable and maybe most importantly reusable. Such a reusability of results helps at eliminating redundancies and thus lowers unnecessary emissions associated with computationally intensive tasks.

A very comprehensive overview of potential energy and carbon emission savings for computing clusters can be found in Refs. \cite{bruers2023resourceawareresearchuniversematter} and \cite{suarez2024energyawareoperationhpcsystems}. In order to operate a computing cluster that complies with increasingly stringent emission regulations in Germany \cite{Bundesgesetzblatt_EnEfG_2023}, progress is needed across all these aspects. Whether by modifying clusters to utilise waste heat, modernising hardware or providing intensive training for end users, progressing towards a carbon-neutral operation inevitably involves considerable effort and expenses.
 	
However, these measures vary significantly in terms of the effort required for implementation. Particularly when targeting short- and medium-term emission reductions, the dynamic operation of a cluster, as introduced in \cref{sec:intro}, can represent a promising alternative. Particularly when assuming that an existing cluster is not operating at full capacity, enabling dynamic operation involves significantly less work than, for example, completely rebuilding a cluster in order to benefit of modern cooling technology and hardware. 

Given that the installed hardware still has spare capacity, a cluster generally needs to be equipped with a control software to implement dynamic operation. Based on publicly available weather data and information on energy generation, this software can automatically decide whether to accept or continue computing tasks at a given time, or whether the cluster should switch to an idle state and resume work when conditions are more favourable.  Such a cut-off in terms of carbon emissions per \unit{MWh} up to which a operation is deemed favourable could offer significant short-term potential in terms of carbon emission and operation cost reduction. In addition, given some flexibility with respect to key parameters, once implemented such a control software could be re-used for different clusters.



\section{Methodology}
\label{sec:methodology}

\begin{figure}[htbp]
	\begin{subfigure}[b]{\textwidth}
		\includegraphics[width=\textwidth]{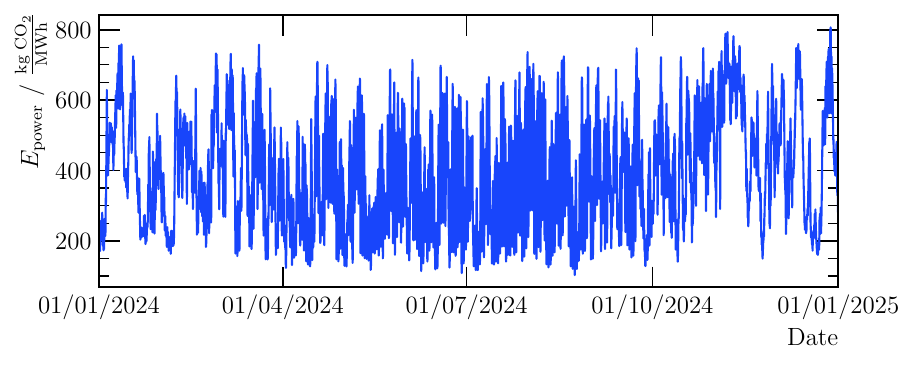}
		\caption{}
		\label{fig:methodology:emission}	
	\end{subfigure}%
	\\
	\begin{subfigure}[b]{\textwidth}
		\includegraphics[width=\textwidth]{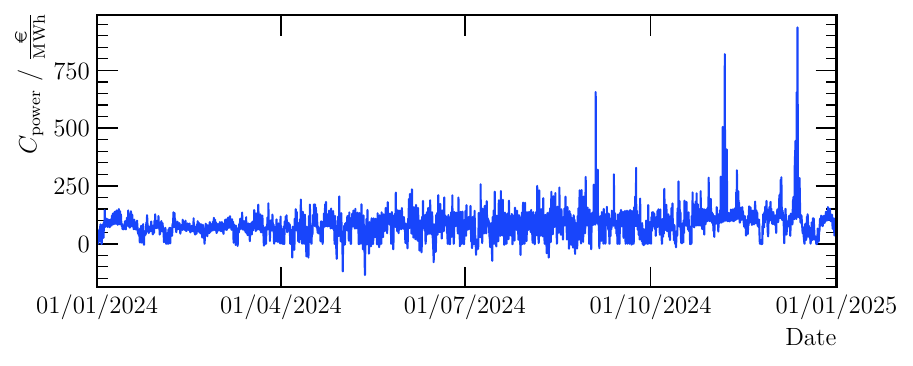}
		\caption{}
		\label{fig:methodology:cost}
	\end{subfigure}%
	\caption{Development of \subref{fig:methodology:emission} carbon emissions in \unit{kg\ce{CO2}\per MWh} and \subref{fig:methodology:cost} costs in \unit{\EUR \per MWh} arising from German net electricity generation for the calendar year 2024.}
	\label{fig:methodology}
\end{figure}

The conducted simulation\footnote{The complete simulation workspace can be found under \url{https://gitlab.uni-bonn.de/grp_phy/drittmittelprojekte/erum-data/susfecit/eco-clusters-study}.} is intended to assess the optimal operation of a computing cluster in the field of high-energy physics. Both the minimisation of carbon emissions and reductions with respect to a computing cluster's operating costs are being targeted.
For the sake of simplicity, it is assumed that the cluster is entirely powered by spot market-traded electricity. The data, covering both the cost of consumed electricity in \unit{\EUR \per MWh} as well as the net electricity generation in \unit{MWh}, was obtained using the available energy data in Germany, publicly provided by the Fraunhofer Institute for Solar Energy Systems ISE\footnote{The website can be found under \url{https://energy-charts.info}.} \cite{FraunhoferISE_EnergyCharts_energy}. Their database collects net electricity production broken down by generation type along with the spot market prices for time intervals of 15 minutes. The corresponding carbon emissions associated with each generation type per year can be accessed separately via ref.\\ \cite{FraunhoferISE_EnergyCharts_emission}, allowing the carbon emissions to be calculated in \unit{kg\ce{CO2}\per MWh} for each 15 minute interval. The resulting price and carbon emission development over a one-year period are shown exemplary for the calendar year 2024 in \cref{fig:methodology:emission} and \cref{fig:methodology:cost}. 

As already described in \cref{sec:background}, a short-term delay when processing computing tasks is acceptable during time periods where electricity is either expensive or associated with high carbon emissions. Nevertheless, the total available computing power for research should remain approximately constant over a longer period of time. This implies that a cluster which, due to high carbon emissions, can only be operated for \qty{50}{\percent} of a given time period must, assuming linear scalability, be designed twice as large. This way a constant compute target can be maintained. For the sake of simplicity, a homogeneous cluster consisting of identical CPUs was simulated in this study. This allows the variable cluster size, depending on the possible cluster utilisation to be modelled directly via the number of logical cores of all combined CPUs. Conversely, the estimated embedded carbon emission for every single logical core has to factor in all components required to expand the computing clusters.

Electricity consumption and thus also the operating emissions for day-to-day operations were calculated based on the computing cluster's estimated electricity consumption and the aforementioned data regarding public net electricity production in Germany. Similar to the embedded carbon emissions, the total power consumption of the cluster was calculated based on the scaling factor, i.e. the number of logical cores. To allow for a more realistic simulation, different workload scenarios were specified.
Each scenario is defined via the respective partial load power consumptions and the relative time fractions during which the computing cluster is operated at each load. This is particularly relevant to avoid neglecting the power consumption of a computing cluster during idle phases. To enable rapid start-up under more favourable conditions and limit hardware damage, complete shutdown was not investigated. Therefore, even in the case of high emissions associated with the net electricity production, i.e.\ periods in which operation is generally avoided, a non-zero electricity consumption was assumed.

Under these assumptions, the total carbon emissions \(E_\text{total}(u)\) are calculated through
\begin{align*}
	E_\text{total}(u)=&\underbrace{\frac{n_\text{cores}}{u} \cdot E_\text{embedded, CPU hour}\cdot t_\text{total}}_{E_\text{embedded}} 
	+ \underbrace{\frac{n_\text{cores}}{u} \cdot P_\text{avg.}\cdot \sum_{i}\left(E_{\text{power},i}\cdot t\right)}_{E_\text{operation}} \\
	&+ \underbrace{\frac{n_\text{cores}}{u} \cdot P_\text{idle}\cdot \sum_{j}\left(E_{\text{power},j}\cdot t \right)}_{E_\text{idle}}\,,
\end{align*} 
where \(n_\text{cores}\) denotes the number of logical cores, \(u\) the relative utilisation within the considered period \(t_\text{total}\), and \(t\) the respective sampling time interval. The carbon emissions of electricity generation in \unit{kg\ce{CO2}\per MWh} for each time interval are represented by \(E_\text{power}\). The index \(i\) runs over all intervals in which \(E_\text{power}\) is below the utilisation threshold value \(X_\text{emission}\). In contrast, the index \(j\) covers the time intervals in which the cluster remains in an idle state as \(E_\text{power}\) exceeds \(X_\text{emission}\). The parameters \(P_\text{avg.}\) and \(P_\text{idle}\) represent the workload scenario dependent average power consumption per logical core during operation phases and idle phases, respectively. An overview of all parameters used in the simulation along with their definition is given in \cref{tab:parameters}.

The optimal threshold value for the emissions associated with the net power generation can be found by minimising \(E_\text{total}(u)\) as a function of the relative cluster utilisation \(u\). This minimisation is achieved numerically by varying the threshold between the minimum and maximum values of \(E_\text{power}\). For each threshold value, the corresponding utilisation and thus ultimately the total emissions \(E_\text{total}(u)\) are derived. Any time the net power production exceeds the optimal threshold associated to the minimum of \(E_\text{total}(u)\) it is more favourable to shut down the computing cluster, i.e. remain in idle mode. A minimum far from \(u=1\) implies, that the computing cluster must be extended to fulfil the constraint arising from the constant compute target. 


Similarly, an optimisation was carried out with regard to the financial costs of operating a computing cluster. As the constant compute target was again modelled via the number of logical cores \(n_\text{cores}\), the price for the purchase of a single core \(C_\text{acq.}\) must reflect all additional acquisition costs for the cluster expansion. Once this price is defined, the total costs \(C_\text{total}(u)\) can be calculated via
\begin{align*}
	C_\text{total}(u)&= 
	\underbrace{\frac{n_\text{cores}}{u}\cdot C_\text{acq., CPU hour}\cdot t_\text{total}}_{C_\text{acq.}} + 
	\underbrace{\frac{n_\text{cores}}{u}\cdot P_\text{max}\cdot C_\text{yearly demand} \cdot t_\text{total}}_{C_\text{demand}}\\
	+&  \underbrace{\frac{n_\text{cores}}{u}\cdot P_\text{avg.}\cdot \sum_{i}\left(C_{\text{power},i}\cdot t\right)}_{C_\text{operation}} +
	\underbrace{\frac{n_\text{cores}}{u} \cdot P_\text{idle}\cdot \sum_{j}\left(C_{\text{power},j}\cdot t\right)}_{C_\text{idle}}\,.
\end{align*} 
Here, \(P_\text{max}\) denotes the maximum power consumption per core, which, when multiplied by the number of cores and the electricity provider’s yearly demand charge, constitutes the second term of the total cost. The operating costs can be determined using the price per power produced at each sampling interval \(C_\text{power}\) and the average power consumption \(P_\text{avg.}\) of a defined workload scenario. Electricity costs during idle periods are determined via the idle power consumption \(P_\text{idle}\). The used indices are defined analogously to the emissions calculation, i.e., \(i\) again covers all intervals in which \(C_\text{power}\) lies below a derived threshold value \(X_\text{cost}\) in \unit{\EUR \per MWh}, while \(j\) runs over all intervals exceeding this threshold.
The base charge of any electricity contract was disregarded in this model, as it would only add an offset to the overall function. The minimum of \(C_\text{total}(u)\) is determined analogously to \(E_\text{total}(u)\).

A definition of all parameters used in the financial cost simulation can be found in \cref{tab:parameters}.

\begin{table}[htbp]
	\caption{Simulation parameters and their definitions.}
	\label{tab:parameters}
	\centering
	\resizebox{\textwidth}{!}{%
	\begin{tabular}{c | c | l }
		\toprule
		Parameter & Unit & Description \\
		\midrule
		\(n_\text{cores}\) & - & Default number of logical cores in a computing cluster.\\
		\(u\) & - & Relative operating time of the cluster throughout the simulated period (\(0\leq u\leq1\)). \\
		\(E_\text{total}(u)\) & \unit{kg\ce{CO2}} & Total carbon emission throughout the simulated period.\\
		\(E_\text{embedded, CPU hour}\) & \unit{kg\ce{CO2}\per h} & Hourly embedded emissions per logical core, based on an assumed lifetime of ten years for all components.\\
		\(E_\text{embedded}\) & \unit{kg\ce{CO2}} & Total embedded emission throughout the simulated period.\\
		\(E_\text{power}\) & \unit{kg\ce{CO2}\per MWh} & Carbon emissions for electricity generated in Germany at each sampled time interval.\\
		\(E_\text{operation}\) & \unit{kg\ce{CO2}} & Total carbon emissions from power consumption during cluster operation throughout the simulated period. \\
		\(E_\text{idle}\) & \unit{kg\ce{CO2}} & Total carbon emissions from power consumption during idling throughout the simulated period.\\ 
		\(C_\text{total}(u)\) & \unit{\EUR} & Total cost throughout the simulated period.\\
		\(C_\text{acq., CPU hour}\) & \unit{\EUR \per h} & Hourly acquisition cost per logical core, based on an assumed lifetime of ten years for all components.\\
		\(C_\text{acq.}\) & \unit{\EUR} & Total acquisition cost throughout the simulated period.\\
		\(C_\text{yearly demand}\) & \unit{\EUR \per kW \per year} & Electricity supplier’s capacity-based charge per year. \\
		\(C_\text{demand}\) & \unit{\EUR} & Total demand cost throughout the simulated period.\\
		\(C_\text{power}\) & \unit{\EUR \per MWh} & Spot market energy price for electricity generated in Germany at each sampled time interval.\\
		\(C_\text{operation}\) & \unit{\EUR} & Total cost from power consumption during cluster operation in the simulated period.\\
		\(C_\text{idle}\) & \unit{\EUR} & Total cost from power consumption during the cluster's idle state in the simulated period.\\
		\(P_\text{avg.}\) & \unit{MW} & Workload-dependent average power consumption per logical core. \\
		\(P_\text{max}\) & \unit{MW} & Maximum power consumption per logical core. \\
		\(P_\text{idle}\) & \unit{MW} & Power consumption per logical core when in idle mode. \\
		\(t_\text{total}\)   & \unit{h}  &  Total time period used in the simulation.\\ 
		\(t\)              & \unit{h}  &  Sampling interval duration. \\ 
		\(X_\text{emission}\) &  \unit{kg\ce{CO2}\per MWh} & Emission threshold above which the computing cluster remains in idle state. \\
    	\(X_\text{cost}\) & \unit{\EUR \per MWh} & Cost threshold above which the computing cluster remains in idle state. \\
		\bottomrule
	\end{tabular}
	}
\end{table}
In order to evaluate the determined emission threshold's stability over a long period, the optimal threshold was extrapolated to a test period covering the calender years 2023 and 2025 for specific scenarios. To allow for a realistic comparison, the extrapolated threshold \(X^\text{extra.}_\text{emission}\) factors in the changed share of renewable energy in Germany. This changed share's impact on the emissions was derived via data on the public energy production's average carbon emission per year and the relative share of renewable energies, again provided by the Fraunhofer ISE platform \cite{FraunhoferISE_EnergyCharts_share}. Assuming a linear relationship over the years from 2002-2025, an increase of \qty{1}{\percent} in the share of renewable energies was found to corresponds to an average emission reduction of \qty{5.4\pm0.3}{kg}. Thus, the extrapolated threshold is given through
\begin{align*}
	X^\text{extra.}_\text{emission} = X_\text{emission}\cdot \qty{5.4\pm0.3}{kg}\cdot(s_{2024}-s_\text{val.})\,,
\end{align*}
where \(s_{2024}\) and \(s_\text{val.}\) correspond to the share of renewable energies in the year 2024 and the chosen validation year, respectively.
In both years, this extrapolated threshold \(X^\text{extra.}_\text{emission}\) was compared to thresholds explicitly derived for each validation year \(X^\text{val.}_\text{emission}\) by again minimising \(E_\text{total}(u)\). 




\section{Simulation results}
\label{sec:result}

The dynamic operation of computing clusters was simulated based on the 2024 data. 
The optimisation with regard to the total carbon emissions, discussed in \cref{sec:results_carbon_emission}, was carried out for the following five different computing clusters setups:
\begin{itemize}
	\item \(\text{BAF}_\text{default}\): Default worker nodes of the Bonn Analysis Facility (BAF) high throughput cluster installed at the Physics Institute at the University of Bonn.
	\item \(\text{BAF}_\text{modern}\): Modern worker nodes of the BAF high throughput cluster installed at the Physics Institute at the University of Bonn.
	\item \(\text{DEEP}_\text{CM}\): Part of the DEEP modular supercomputer prototype \cite{eicker2016deep} at the Jülich Supercomputing Centre (JSC) \cite{jsc}. Designed for high single-thread performance and a modest amount of memory.  
	\item \(\text{DEEP}_\text{DAM}\): Part of the DEEP modular supercomputer prototype at the JSC. Designed for data-intensive analytics and machine learning applications. Accelerators are ignored in the simulation.
	\item \(\text{GridKa}_\text{ARM}\): ARM worker nodes in the WLCG’s Tier 1 Grid Computing Centre Karlsruhe (GridKa) \cite{gridka}.
\end{itemize}
It is important to note, that the tested cluster setups do not exactly correspond to their real-life counterparts. Instead, a simplified setup, neglecting heterogeneous hardware configurations and accelerators was simulated.
The optimisation in terms of total operating costs, discussed in \cref{sec:results_operating_costs}, was only carried out for the two BAF cluster scenarios.

\subsection{Optimisation with respect to carbon emission}
\label{sec:results_carbon_emission}
An overview of the parameters used for emissions optimisation across all the different scenarios can be found in \cref{tab:setup_parameters}. The first difference between the setups concerns the maximum power consumption per logical core \(P_\text{max}\), which was determined via full load tests. The resulting values naturally depend on the hardware used as well as the specific cluster settings, and vary between \qty{9.2}{W} (\(\text{BAF}_\text{default}\)) and \qty{2.9}{W} (\(\text{GridKa}_\text{ARM WN}\)). The associated idle values \(P_\text{idle}\) range between \qty{15}{\percent} (\(\text{BAF}_\text{modern}\)) and \qty{48}{\percent} (\(\text{DEEP}_\text{DAM}\)) of the respective maximum power consumption.  

\begin{table}[htbp]
	\caption{Parameter values chosen for the simulation of all five cluster scenarios.}
	\label{tab:setup_parameters}
	\centering
	\resizebox{\textwidth}{!}{%
	\begin{tabular}{c | c | c | c | c | c }
		\toprule
		\backslashbox{Parameter}{Setup} & \(\text{BAF}_\text{default}\) & \(\text{BAF}_\text{modern}\) & \(\text{DEEP}_\text{CM}\) & \(\text{DEEP}_\text{DAM}\) & \(\text{GridKa}_\text{ARM WN}\) \\
		\midrule
		\(n_\text{cores}\) & 7104 & 2816 & 2400 & 1536 & 2816 \\
		\(P_\text{max}\) / \unit{W} & \num{9.2} & \num{7.6} & \num{7.8} & \num{6.9} & \num{2.9} \\
		\(P_\text{idle}\) / \unit{W} & \num{2.3} & \num{1.1} & \num{2.6} & \num{3.3} & \num{0.9} \\
		\(E_\text{embedded}\) / \unit{kg\ce{CO2} \per h} & \num{1.5e-5} & \num{1.8e-4} & \num{1.8e-4} & \num{1.8e-4} & \num{1.8e-4} \\
		\bottomrule
	\end{tabular}
	}
\end{table}
Whilst the power consumption of each individual setup was directly measured, no precise data was available for the embedded emissions of the installed hardware. Therefore, \(E_\text{embedded, CPU hour}\) in all five setups was estimated using a life cycle assessment of the Dell R740 rack server conducted by the thinkstep AG \cite{Dell_LCA_R740_2019}. The study lists embedded emissions of \qty{4092}{kg\ce{CO2}} per rack server, taking into account manufacturing costs (\qty{4288}{kg\ce{CO2}}), transport costs (\qty{3}{kg\ce{CO2}}) and end of life credits (\qty{-199}{kg\ce{CO2}}). With approximately \qty{80.3}{\percent} the dominant share of manufacturing emissions is accounted for by the SSD production. To express the embedded emissions per logical core, it was assumed that \num{256} logical cores can be installed per rack server. Assuming a life cycle of 10 years, this resulted in \(E_\text{embedded, CPU hour}=\qty{1.8e-4}{kg\ce{CO2} \per h}\) for all but the \(\text{BAF}_\text{default}\) setup.

In contrast to the others, the \(\text{BAF}_\text{default}\) setup, solely relies on HDDs as storage media, for which significantly lower embedded emissions had to be assumed. As a reference, the study conducted in \cite{Tannu_2023}, which compared life cycle assessments from various manufacturers to derive an average storage emission factor \(\text{SEF}\) of \qty{0.16}{kg\ce{CO2} \per GB} for SSDs and \qty{0.02}{kg\ce{CO2} \per GB} for HDDs, was used.  

Applying these storage emission factors, the \(\text{BAF}_\text{default}\) setup's embedded emissions amounted to \(\qty{1.5e-5}{kg\ce{CO2} \per h}\). It should be noted at this point that the embedded emissions evaluated in ref.\ \cite{Tannu_2023} displayed significant fluctuations, especially when comparing different manufacturers. Thus, the exact value of \(E_\text{embedded}\) is highly dependent on the chosen setup and the installed hardware. All values chosen here are only rough estimations and assumed to be associated with considerable uncertainties. 

In order to simulate realistic use cases of a computing cluster, different workload scenarios were considered for all setups. Besides a medium workload and a heavy workload scenario, which are oriented towards studies in ref.\ \cite{Dell_LCA_R740_2019}, a backfilling \cite{SHMUELI20051090} scenario was defined. Such a scenario is frequently used in high-energy physics and allows free resources to be allocated more efficiently by prioritising short single-core jobs. As a result, the cluster can be assumed to run under full load during the simulated period. The exact parameters of all three scenarios are summarised in \cref{tab:workloads}. For each load working point, it was assumed that a cluster's power consumption scales linearly between \(P_\text{idle}\) and \(P_\text{max}\).

\begin{figure}[htbp]
	\includegraphics[width=\textwidth]{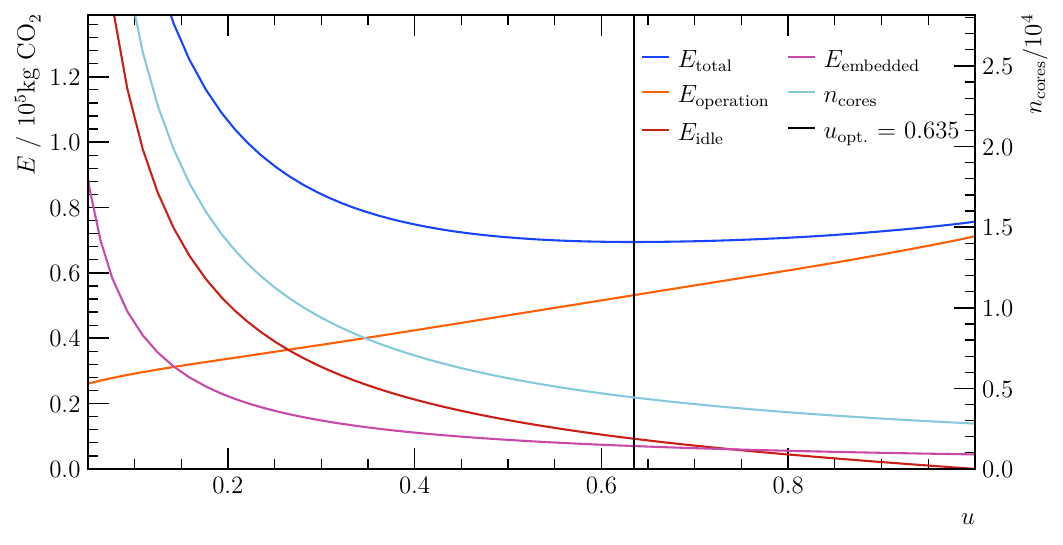}
	\caption{Display of total emission alongside the individual contributions (left \(y\)-axis) and number of logical cores needed to maintain the default compute target (right \(y\)-axis) as a function of the utilisation \(u\) for the \(\text{BAF}_\text{modern}\)'s backfilling scenario. The optimal cluster utilisation is marked by the black vertical line. Whilst \(E_\text{operation}\) increases with rising utilisation, \(E_\text{idle}\) and \(E_\text{embedded}\) decrease. The latter results from reduced hardware requirements to meet the constant computing target.}
	\label{fig:results_carbon_emission_baf_modern}
\end{figure}

\begin{table}[htbp]
	\caption{Workload scenarios used in the simulation. Setup dependent power consumption is assumed to scale linearly between \(P_\text{idle}\) and \(P_\text{max}\).}
	\label{tab:workloads}
	\centering
	\begin{tabular}{c | c | c | c | c}
		\toprule
		Load mode / \unit{\percent} & 0 & 10 & 50 & 100 \\
		\midrule
		Medium workload time fraction / \unit{\percent} & 25 & 30 & 35 & 10 \\
		Heavy workload time fraction / \unit{\percent} & 10 & 20 & 55 & 15 \\
		Backfilling workload time fraction / \unit{\percent} & 5 & 0 & 0 & 95 \\
		\bottomrule
	\end{tabular}
\end{table}
The optimisation results of the five setups are summarised in \cref{tab:results_carbon_emission} for all workload scenarios. In addition, \(E_\text{total}(u)\) alongside the individual contributions is displayed in \cref{fig:results_carbon_emission_baf_modern} as a function of the utilisation \(u\) in case of \(\text{BAF}_\text{modern}\)'s backfilling scenario and illustrates the expected trend: \(E_\text{operation}\) decreases along with the utilisation, as the cluster uses electricity with increasingly lower levels of associated carbon emissions. At the same time, \(E_\text{idle}\) and \(E_\text{embedded}\) along with the number of logical cores, i.e. the cluster scaling factor, increase to maintain a constant compute target. 

\begin{table}[htbp]
	\caption{Optimisation results to minimise carbon emissions for all setups and workload scenarios. Listed is the optimal utilisation \(u_\text{opt.}\), the associated threshold \(X_\text{emission}\) above which a computing cluster switches to idle state, and the relative emissions compared to constant operation \(\nicefrac{E_\text{total}(u_\text{opt.})}{E_\text{total}(u=1)}\). In case that a constant operation yields the lowest total carbon emissions, no threshold on \(E_\text{power}\) is specified.}
	\label{tab:results_carbon_emission}
	\centering
		\begin{tabular}{c | c | c | c}
			\toprule
			Workload scenario & \(u_\text{opt.}\) & \(X_\text{emission}\) / \unit{kg\ce{CO2}\per MWh} & \(\nicefrac{E_\text{total}(u_\text{opt.})}{E_\text{total}(u=1)}\) \\
			\midrule
			\multicolumn{4}{c}{\(\text{BAF}_\text{default}\)} \\
			\midrule
			Medium & 1.000 & - & 1.000 \\
			Heavy & 0.975 & 699.90 & 0.998\\
			Backfilling & 0.745 & 515.01 & 0.955 \\
			\midrule
			\multicolumn{4}{c}{\(\text{BAF}_\text{modern}\)} \\
			\midrule
			Medium & 0.983 & 714.13 & 0.999 \\
			Heavy & 0.882 & 607.45 & 0.988\\
			Backfilling & 0.635 & 458.11 & 0.918 \\
			\midrule
			\multicolumn{4}{c}{\(\text{DEEP}_\text{CM}\)} \\
			\midrule
			Medium & 1.000 & - & 1.000 \\
			Heavy & 1.000 & - & 1.000 \\
			Backfilling & 0.916 & 635.90 & 0.993 \\
			\midrule
			\multicolumn{4}{c}{\(\text{DEEP}_\text{DAM}\)} \\
			\midrule
			Medium & 1.000 & - & 1.000 \\
			Heavy & 1.000 & - & 1.000 \\
			Backfilling & 1.000 & - & 1.000 \\
			\midrule
			\multicolumn{4}{c}{\(\text{GridKa}_\text{ARM}\)} \\
			\midrule
			Medium & 1.000 & - & 1.000 \\
			Heavy & 1.000 & - & 1.000 \\
			Backfilling & 0.953 & 671.458 & 0.997 \\
			\bottomrule
		\end{tabular}
\end{table}

A comparison across the five setups generally demonstrates that emission decrease through dynamic operation is more pronounced for workloads with high cluster utilisation. In addition, it is apparent that the emission reduction strongly depends on the ratio between \(P_\text{idle}\) and \(P_\text{max}\). If power consumption is not reduced significantly when idling, the benefits of a dynamic computing cluster operation are negated. Consequently, the greatest potential for savings is found for the \(\text{BAF}_\text{modern}\) setup when simulated for the backfilling workload scenario. At a utilisation of \(u=0.635\), carbon emissions can be reduced by around \qty{8}{\percent} compared to \(u=1\). In contrast, the \(\text{DEEP}_\text{DAM}\) setup offers no potential for savings. Its relatively high power consumption in idle mode (\(\nicefrac{P_\text{idle}}{P_\text{max}}\approx\qty{48}{\percent}\)) results in a trivial minimum of \(u=1\).

\begin{figure}[htbp]
	\includegraphics[width=\textwidth]{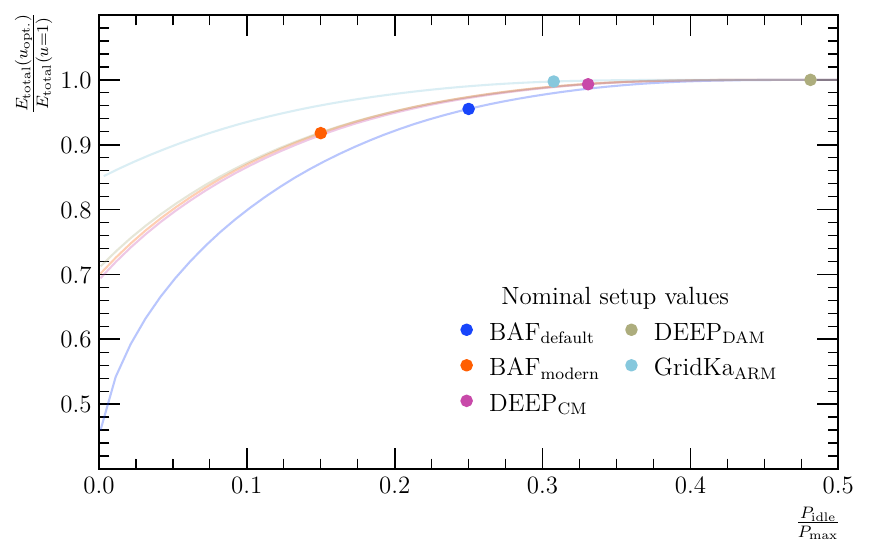}
	\caption{Relative emissions compared to constant operation \(\nicefrac{E_\text{total}(u_\text{opt.})}{E_\text{total}(u=1)}\) as a function of the power consumption ratio \(\nicefrac{P_\text{idle}}{P_\text{max}}\) for all setups in the backfilling scenario. Nominal power consumption ratios are marked.}
	\label{fig:overview_plot_idle_power}
\end{figure}

The actual idle power consumption is not only hardware-dependent but can also be adjusted by altering hardware configuration\footnote{For example, BIOS energy settings.}. To illustrate potential variability and assess its effect, \cref{fig:overview_plot_idle_power} shows the relative emission reduction as a function of the relative idle power consumption \(\nicefrac{P_\text{idle}}{P_\text{max}}\). For better visualisation, only the backfilling scenario is displayed. 

The strong dependency of the emission reduction on the ratio is evident. Potential savings compared to a continuous operation decrease rapidly as the relative idle power consumption increases. Beyond a threshold of \(\nicefrac{P_\text{idle}}{P_\text{max}}\approx 0.4\), dynamic operation no longer offers any advantages regardless of the simulated setup. In contrast, assuming a cluster can be completely shut down instead of idling (\(\nicefrac{P_\text{idle}}{P_\text{max}}=0\)), savings of more than \qty{50}{\percent} could be achieved depending on the setup.

To further estimate which effect the uncertainty with regard to embedded emissions has on the overall optimisation, the relative savings were derived as a function of embedded emissions. This test was only performed for the \(\text{BAF}_\text{modern}\)'s backfilling scenario. All other simulation parameters were kept unchanged. The results for an up- and down-variation of \qty{50}{\percent} with respect to the nominal value of \qty{1.8e-4}{kg\ce{CO2} \per h} are listed in \cref{tab:results_carbon_emission_variable_embedded}.
The results show the expected trend: increased embedded emissions shift the minimum closer to \(u=1\) and vice versa. Even for the up-variation, the calculated optimum corresponds to a utilisation smaller than \(u=1\). The associated relative emission reduction lies at approximately \qty{5}{\percent}. In the tested setup, dynamic operation of a cluster enables a carbon emission reduction across a fairly wide range of \(E_\text{embedded}\). This is consistent with the results shown in \cref{fig:results_carbon_emission_baf_modern} which already illustrated that the embedded emissions do not represent the optimisation's dominant factor.
\begin{table}[htbp]
	\caption{Optimisation results to minimise carbon emissions for the \(\text{BAF}_\text{modern}\) setup's backfilling scenarios. Listed is the optimal utilisation \(u_\text{opt.}\), the associated threshold \(X_\text{emission}\) beyond which a computing cluster switches to idle state, and the relative emissions compared to constant operation \(\nicefrac{E_\text{total}(u_\text{opt.})}{E_\text{total}(u=1)}\). In addition to the nominal value, the results are shown for a \qty{50}{\percent} up- and down-variation of the embedded emissions.}
	\label{tab:results_carbon_emission_variable_embedded}
	\centering
	\begin{tabular}{c | c | c | c}
		\toprule
		Workload scenario & \(u_\text{opt.}\) & \(X_\text{emission}\) / \unit{kg\ce{CO2}\per MWh} & \(\nicefrac{E_\text{total}(u_\text{opt.})}{E_\text{total}(u=1)}\) \\
		\midrule
		\multicolumn{4}{c}{\(\text{BAF}_\text{modern}\)} \\
		\midrule
		Backfilling & 0.635 & 458.11 & 0.918 \\
		Backfilling (\(E_\text{embedded}\times0.5\)) & 0.594 &  436.78 & 0.896 \\
		Backfilling (\(E_\text{embedded}\times1.5\)) & 0.723 & 500.78 & 0.949 \\
		\bottomrule
	\end{tabular}
\end{table}

\begin{figure}[htbp]
	\includegraphics[width=\textwidth]{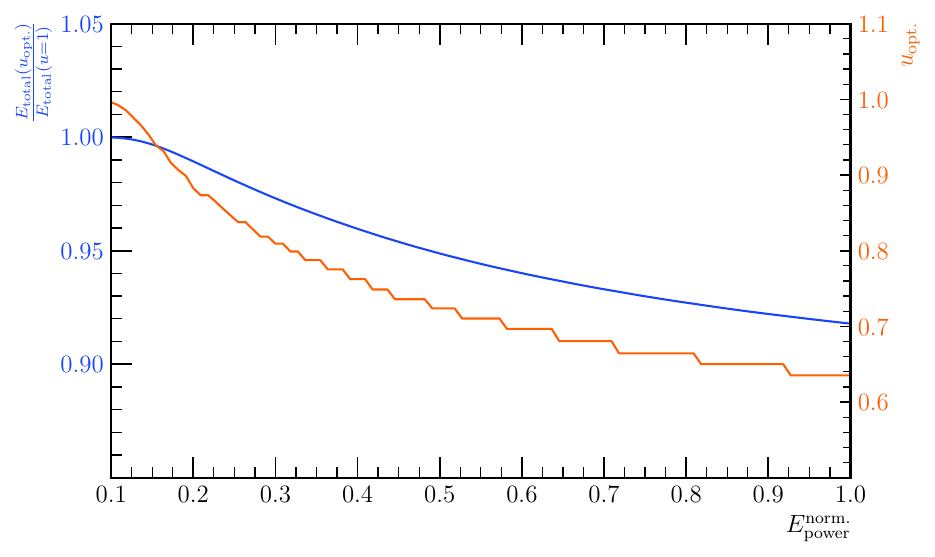}
	\caption{Relative emissions compared to constant operation \(\nicefrac{E_\text{total}(u_\text{opt.})}{E_\text{total}(u=1)}\) (left \(y\)-axis) and optimal utilisation \(u_\text{opt.}\) (right \(y\)-axis) as a function of the normalised carbon emission associated to the electricity generated in Germany \(E^\text{norm}_\text{total}\). Here \(E^\text{norm}_\text{total}=1.0\) corresponds to the actual emission in 2024 while \(E^\text{norm}_\text{total}=0.1\) corresponds to a reduction by \qty{90}{\percent}.}
	\label{fig:overview_plot_emission}
\end{figure}

Lastly, a study was conducted to estimate how an increase of renewable energy sources, i.e.\ lower carbon emissions associated to electricity generation, would affect the saving potential of a dynamic computing cluster operation. As with the variation in embedded emissions, this test was carried out only for the \(\text{BAF}_\text{modern}\)'s backfilling scenario. To cover a wide range, \(E_\text{power}\) was reduced by up to \qty{90}{\percent} in the simulation. The results regarding the relative emission reduction and the corresponding optimal utilisation \(u_\text{opt.}\) are shown in \cref{fig:overview_plot_emission}. As expected, reducing \(E_\text{power}\) causes \(u_\text{opt.}\) to increase and consequently the relative emissions \(\nicefrac{E_\text{total}(u_\text{opt.})}{E_\text{total}(u=1)}\) approximate \num{1}. However, even when reducing \(E_\text{power}\) down to \qty{20}{\percent} of its original value (\(E^\text{norm}_\text{total}=0.2\)), dynamic operation still allows for emission reduction. Thus, the dynamic operation of a computing cluster continues to offer advantages even when assuming way higher shares of renewable energy sources.

\subsubsection{Dynamic operation and clock frequency limitation - comparison}

The \(\text{GridKa}_\text{ARM}\) setup allows for an additional comparison with an alternative cluster operation mode, based on studies conducted in ref.\ \cite{krull2025experience} regarding the installed hardware's power consumption. In this study, the CPUs clock frequency was scanned to find the maximum power efficiency. The latter is defined as the ratio of CPU performance to the average power consumption. In this case, CPU performance was quantified using the HS23 score\footnote{HS23 is a standard benchmark performance unit to measure computing resources uniformly across the Worldwide LHC Computing Grid (WLCG) \cite{WLCG}. It measures performance using typical high-energy physics workloads from multiple experiments.} \cite{szczepanek2024hepbenchmarksuiteenhancing}. At the maximum derived in ref.\ \cite{krull2025experience}, power consumption was reduced by around \qty{40}{\percent}, whilst the HS23 score only dropped by \qty{19}{\percent}.

\begin{figure}[htbp]
	\includegraphics[width=\textwidth]{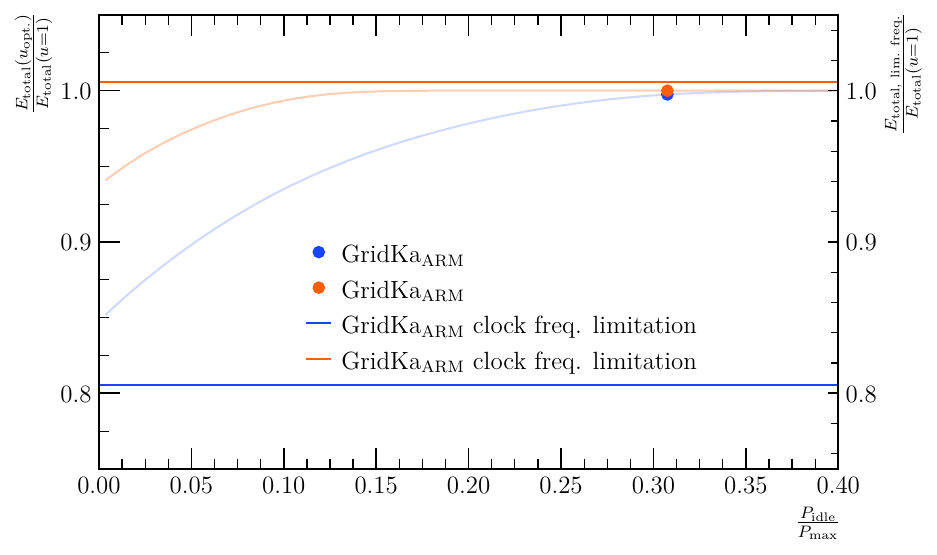}
	\caption{Relative emissions compared to constant operation \(\nicefrac{E_\text{total}(u_\text{opt.})}{E_\text{total}(u=1)}\) (left \(y\)-axis) as a function of the power consumption ratio \(\nicefrac{P_\text{idle}}{P_\text{max}}\) for the \(\text{GridKa}_\text{ARM}\) setup. The nominal power consumption ratio is marked in blue (orange) for the backfilling (medium) scenario. The blue (orange) horizontal line indicates the relative emissions \(\nicefrac{E_\text{total, lim. freq.}}{E_\text{total}(u=1)}\) (right \(y\)-axis) for the backfilling (medium) scenario, assuming continuous utilisation of the cluster at a limited clock frequency. Due to the continuous utilisation varying the power consumption ratio \(\nicefrac{P_\text{idle}}{P_\text{max}}\) has no effect for those scenarios.}
	\label{fig:overview_comparison_plot_gridka}
\end{figure}

If the cluster’s CPUs are permanently limited to the associated frequency, carbon emissions can be reduced similarly to the dynamic operation. To still meet the requirement of a constant compute target, the \qty{19}{\percent} drop in the HS23 score was compensated by additional hardware and the accompanying increase in embedded emissions. Total emission for operation with a limited frequency are denoted as \(E_\text{total, lim. freq.}\). The results of dynamic operation and operation with limited CPU frequency are compared in \cref{fig:overview_comparison_plot_gridka}. Similar to \cref{fig:overview_plot_idle_power}, the nominal simulation is again displayed besides dynamic operation results for a varied power consumption ratio \(\nicefrac{P_\text{idle}}{P_\text{max}}\). A comparison of the results shows that limiting the clock frequency at the medium workload leads to increased emissions caused by the additional hardware. The dynamic operation reaches its optimum at the trivial case of \(u=1\). 
For the backfilling scenario, however, limiting the clock frequency results in an emissions reduction of almost \qty{20}{\percent}, thereby significantly outperforming dynamic operation of the \(\text{GridKa}_\text{ARM}\) setup regardless of the power consumption ratio.

Thus, the feasibility of such a clock frequency limit is, not only highly dependent on the hardware but also affected by the cluster's typical workloads.
However, given a clear power efficiency maximum is found, a clock frequency limit could be combined with dynamic operation of a computing cluster. A simulation of this combination is not part of this study but offers an interesting additional test case.

\subsubsection{Emission operation threshold - validation}
\label{sec:results:validation}

In order to verify the emission minimisation results, in particular the derived threshold's stability, a validation was carried out based on the 2023 and 2025 data. As an example, the \(\text{BAF}_\text{modern}\) setup's backfilling scenario was used. The extrapolation defined in \cref{sec:methodology} was applied. A comparison between the targeted utilisation \(u_\text{target}\), associated to the threshold \(X_\text{emission}\) determined via the 2024 data, and the utilisation \(u_\text{extra.}\) reached with the extrapolated thresholds \(X^\text{extra.}_\text{emission}\) is listed in \cref{tab:validation}. Resulting deviations are within two percent, which implies that the compute target is almost perfectly met in both validation periods.

\begin{figure}[htbp]
	\begin{subfigure}[b]{\textwidth}
		\includegraphics[width=\textwidth]{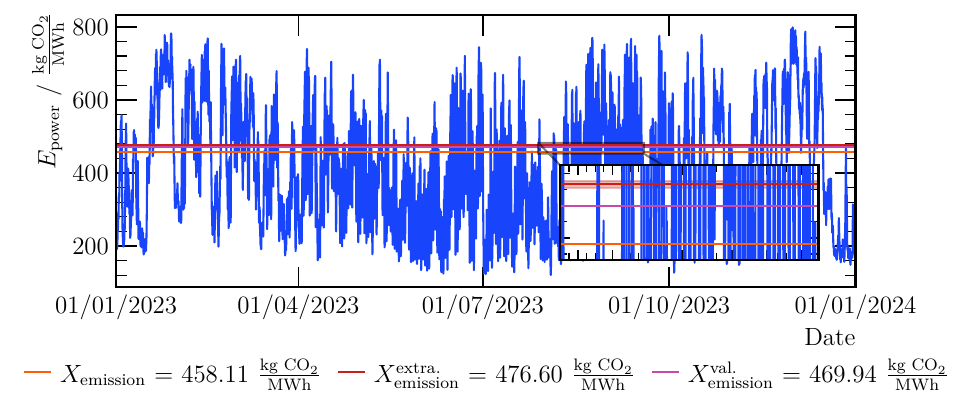}
		\caption{}
		\label{fig:validation:past}
	\end{subfigure}%
	\\
	\begin{subfigure}[b]{\textwidth}
		\includegraphics[width=\textwidth]{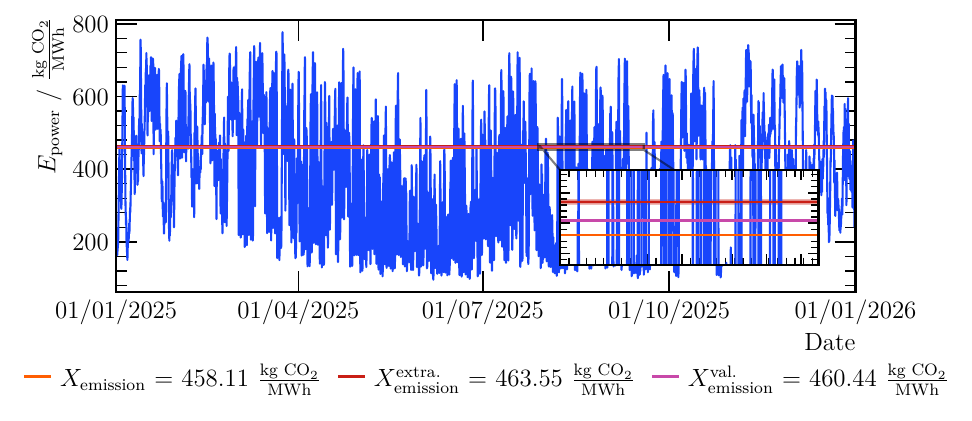}
		\caption{}
		\label{fig:validation:future}
	\end{subfigure}%
	\caption{Validation of the calculated threshold for the \(\text{BAF}_\text{modern}\)'s backfilling scenario in the years \subref{fig:validation:past} 2023 and \subref{fig:validation:future} 2025. Both validation periods display the original threshold value \(X_\text{emission}\) determined via the 2024 data, the extrapolated threshold \(X^\text{extra.}_\text{emission}\) taking into account the changed share of renewable energies as well as the threshold \(X^\text{target}_\text{emission}\) derived individually in each validation period to match \(u_\text{target}\).}
	\label{fig:validation}
\end{figure}

The extrapolated thresholds \(X^\text{extra.}_\text{emission}\) are displayed in \cref{fig:validation}. In addition both are compared to the threshold \(X^\text{target}_\text{emission}\), individually derived to match \(u_\text{target}\) in each validation period. This comparison yields relative deviations of \qty{1.42}{\percent} and \qty{0.68}{\percent} for the years 2023 and 2025, respectively. Therefore, taking into account the changed share of renewable energies, no significant deviations are found within both validation periods. 

\begin{table}[htbp]
	\caption{Target utilisation \(u_\text{target}\) derived via the 2024 data and utilisation obtained in a validation period \(u_\text{extra.}\) taking into account the changed share of renewable energies for the \(\text{BAF}_\text{modern}\) setup's backfilling scenario. Associated emission thresholds are also listed.}
	\label{tab:validation}
	\centering
	\begin{tabular}{c | c  c | c  c}
		\toprule
		Validation period & \(u_\text{extra.}\) & \(X^\text{extra.}_\text{emission}\) / \unit{kg\ce{CO2}\per MWh} & \(u_\text{target}\) & \(X^\text{target}_\text{emission}\) / \unit{kg\ce{CO2}\per MWh} \\
		\midrule
		2023  & 0.648 & 476.60 & 0.635 & 469.94 \\
		2025  & 0.648 & 463.55 & 0.635 & 460.44 \\
		\bottomrule
	\end{tabular}
\end{table}

\subsection{Optimisation with respect to operational costs}
\label{sec:results_operating_costs}

Due to a lack of data, optimisation with respect to operating costs was solely carried out for the two different BAF computing cluster setups. 
Both setups were simulated under the assumption that \qty{100}{\percent} of the electricity demand is purchased on the spot market. A value of \(C_\text{demand}=\qty{100}{\EUR \per kW \per year}\) was assumed for the demand charge. This was again only intended to serve as a rough estimate, as grid fees in Germany generally depend on the supplier, the region and the duration of use. 
The acquisition costs per logical core hour were assumed to lie at \(C_\text{acq., CPU hour}=\qty{5.35e-4}{\EUR \per h} \) for both scenarios. This value was determined based on the costs of BAF hardware currently in use. The calculation only incorporates purchase prices for CPUs and RAM while additional costs for infrastructure etc. are excluded. An internal comparison between original acquisition costs in early 2017 and the costs for more recent expansions in 2022 and 2023 revealed that the acquisition cost has remained approximately constant to date. It is difficult to predict how acquisition costs will change in the future. However, with global developments in the field of artificial intelligence and the associated demand for RAM modules \cite{zhang2025minutessecondsredefiningfiveminute} a cost increase seems increasingly likely. 

The assumed values used for both setups in terms of power consumption, i.e. \(P_\text{max}\) and \(P_\text{idle}\), still correspond to those listed in \cref{tab:parameters}. Similarly, all three workload scenarios from \cref{tab:workloads} were also reused.

\begin{figure}[htbp]
	\includegraphics[width=\textwidth]{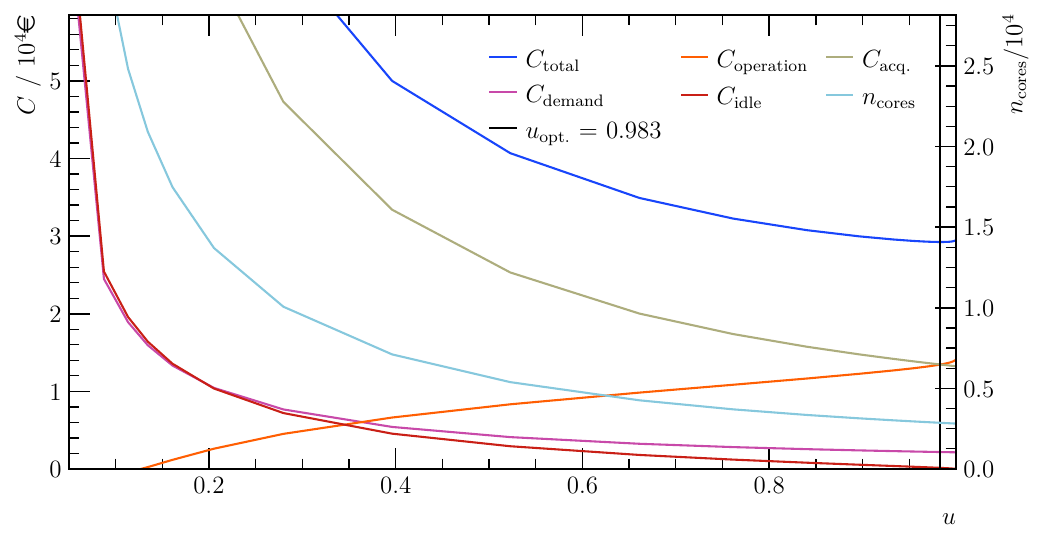}
	\caption{Display of total costs alongside the individual contributions (left \(y\)-axis) and number of logical cores needed to maintain the default compute target (right \(y\)-axis) as a function of the utilisation \(u\) for the \(\text{BAF}_\text{modern}\) setup's backfilling scenario. The optimal cluster utilisation is marked by the black vertical line. Whilst \(C_\text{operation}\) increases with rising utilisation, \(C_\text{idle}\), \(C_\text{demand}\) and \(C_\text{acq.}\) decrease. The latter two results from reduced hardware requirements to meet the constant computing target.}
	\label{fig:results_operation_cost_baf_modern}
\end{figure}

Cost minimisation results are listed in \cref{tab:results_operation_cost} for both setups. Similar to the results in \cref{sec:results_carbon_emission}, \cref{fig:results_operation_cost_baf_modern} additionally illustrates total and individual costs as a function of the utilisation \(u\) in case of \(\text{BAF}_\text{modern}\)'s backfilling scenario. As expected, a decreased utilisation causes not only an increase in \(C_\text{idle}\), but also in \(C_\text{demand}\) and \(C_\text{acq.}\). The latter both represent costs connected to the additional hardware, required to maintain a constant compute target. 

A comparison of all listed scenarios demonstrates that the greatest cost-saving is achieved from simulating a dynamic operation under heavy cluster utilisation, as was also the case for the carbon emission studies. Nevertheless, the relative savings for all tested scenarios lie below one percent. This limited saving potential can be attributed to the dominating acquisition costs \(C_\text{acq.}\), as shown in \cref{fig:results_operation_cost_baf_modern}. They rise for utilisations of \(u<1\), as more hardware is required to keep the compute target constant. 

\begin{figure}[htbp]
	\includegraphics[width=\textwidth]{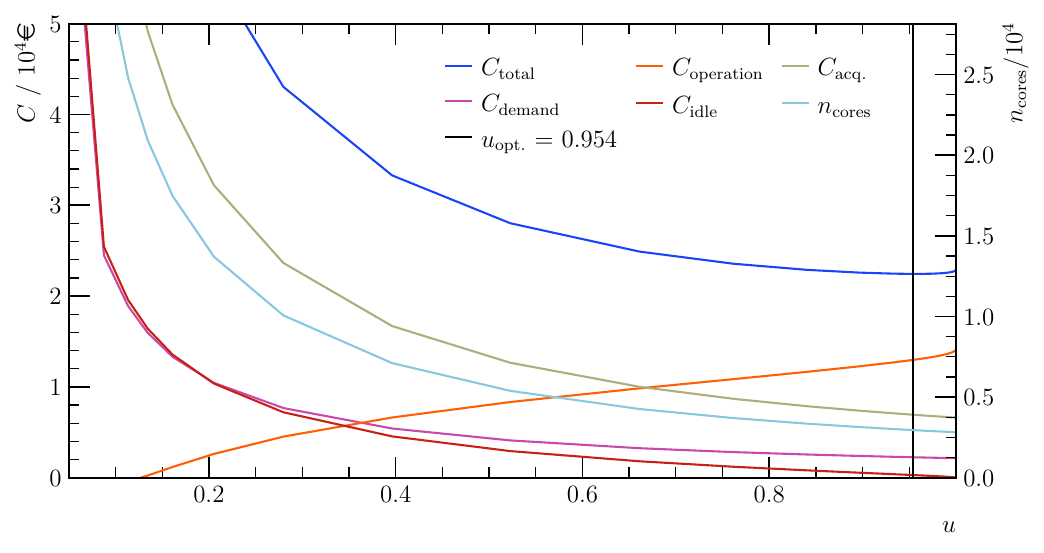}
	\caption{Display of total costs alongside the individual contributions (left \(y\)-axis) and number of logical cores needed to maintain the default compute target (right \(y\)-axis) as a function of the utilisation \(u\) for the \(\text{BAF}_\text{modern}\) setup's backfilling scenario assuming \(C_\text{acq.}\) is reduced by \qty{50}{\percent}. The optimal cluster utilisation is marked by the black vertical line.}
	\label{fig:results_operation_cost_baf_modern_reduced_acq}
\end{figure}

However, the actual acquisition costs for different setups' hardware might vary significantly. To assess the potential of a dynamic operation as a function of the acquisition costs, further simulations were carried out for the \(\text{BAF}_\text{modern}\)'s backfilling scenario. Results for an acquisition cost reduction of \qty{50}{\percent} are shown in \cref{fig:results_operation_cost_baf_modern_reduced_acq}. While \(C_\text{acq.}\) is no longer the dominant contribution for \(u\gtrapprox 0.7\), the relative savings compared to \(C_\text{total}(u=1)\) are still below \qty{2}{\percent}. Consequently, a dynamic operation can, even under the assumption of reduced acquisition costs, only achieve quite limited savings. 

Following the approach in \cref{sec:results_carbon_emission} on emissions, the cost reduction potential was also evaluated across varying electricity prices. To this end, simulations up to a reduction of \(C_\text{power}\) to \qty{10}{\percent} of its original value were carried out for the \(\text{BAF}_\text{modern}\)'s backfilling scenario. In principle, lower electricity prices naturally reduce the savings from a dynamic operation, i.e.\ \(u_\text{opt.}\) increases. The simulation showed that small savings of less than \qty{1}{\percent} are achievable as long as the electricity price \(C_\text{power}\) does not drop by more than \qty{85}{\percent}. Thus, apart from the fact that operational cost savings showed to be quite limited across all tested scenarios, only a significantly reduced electricity price leads to the trivial result of \(u_\text{opt.}=1\).

\begin{table}[htbp]
	\caption{Optimisation results to minimise operational costs for both BAF setups and workload scenarios. Listed is the optimal utilisation \(u_\text{opt.}\), the associated threshold \(X_\text{cost}\) beyond which a computing cluster switches to idle state, and the relative cost compared to constant operation \(\nicefrac{C_\text{total}(u_\text{opt.})}{C_\text{total}(u=1)}\). In case that a constant operation yields the lowest total costs, no threshold on \(C_\text{power}\) is specified.}
	\label{tab:results_operation_cost}
	\centering
	\begin{tabular}{c | c | c | c}
		\toprule
		Workload scenario & \(u_\text{opt.}\) & \(X_\text{cost}\) / \unit{\EUR \per MWh} & \(\nicefrac{C_\text{total}(u_\text{opt.})}{C_\text{total}(u=1)}\) \\
		\midrule
		\multicolumn{4}{c}{\(\text{BAF}_\text{default}\)} \\
		\midrule
		Medium & 0.999 & 470.78 & 0.999\\
		Heavy & 0.996 & 351.70 & 0.998 \\
		Backfilling& 0.985 & 200.14 & 0.992 \\
		\midrule
		\multicolumn{4}{c}{\(\text{BAF}_\text{modern}\)} \\
		\midrule
		Medium & 0.998 & 438.30 & 0.999 \\
		Heavy & 0.996 & 330.05 & 0.998\\
		Backfilling & 0.983 & 189.32 & 0.991 \\
		\bottomrule
	\end{tabular}
\end{table}

\section{Conclusion}
\label{sec:conclusion}
Germany's growth in renewable energies is inevitably linked to relatively large fluctuations with respect to the total energy production. Large power consumers whose operation can be dynamically controlled are able to partially compensate these fluctuations. Through trading on the electricity spot market such a dynamic operation offers potential savings in carbon emission and operating costs. This study investigated the feasibility of dynamically operating a computing cluster used in science. These computing clusters represent an ideal test case, since short-term delays caused by any dynamic operation are often negligible as long as the overall compute target remains constant over prolonged periods. 

On basis of the German energy generation data of 2024, dynamic operation for five different computing clusters installed at research institutes was simulated. The  results were compared with respect to a potential carbon emission and operational cost reduction. Savings with respect to the operating costs during the simulation period were found to be quite limited due to the dominant acquisition costs. For all tested setups and workloads relative savings stayed below \qty{1}{\percent} with respect to a standard non-dynamic operation. 

In contrast, the results regarding carbon emissions showed more pronounced differences between the various cluster configurations. The decisive factor for the effective implementation of dynamic operation was found to be the relative idle power consumption \(\nicefrac{P_\text{idle}}{P_\text{max}}\). Setups with a comparatively high value allowed for little to no emission reduction, whereas the \(\text{BAF}_\text{modern}\) setup with \(\nicefrac{P_\text{idle}}{P_\text{max}}=0.15\) achieved savings of up to \qty{8}{\percent}. 

To validate the results, dynamic operation of the \(\text{BAF}_\text{modern}\) setup's backfilling scenario was additionally simulated for the years 2023 and 2025. Taking into account the changed share of renewable energies, a deviation from the targeted relative utilization of no more than \qty{2}{\percent} was attainable. This demonstrated, at least within the limited validation period, that a fixed annual compute target can be met consistently.

Apart from the two specific setups, simulations were also carried out for variable modelling parameters. Values for relative idle power consumption and acquisition costs were varied independently in order to assess the effects of potential future developments or different hardware setups. To at least partially evaluate the effect of the embedded emissions' large uncertainties additional testes were carried out. It was demonstrated that the dynamic operation allowed for an emission reduction within the chosen uncertainty band. 

The \(\text{GridKa}_\text{ARM}\) setup enabled a comparison between dynamic operation and continuous operation at a limited clock frequency. Results showed to be quite hardware- and workload-dependent but combining both methods might be a viable option to decrease the emission in the future.
 
While this study is intended to serve as a proof of concept, demonstrating the potential for savings, particularly in terms of carbon emissions, further steps are necessary before implementing the dynamic operation of a cluster in practice. 
Among other simplifications used in the simulation, an interface to stop and restart the cluster operation is mandatory. Computing jobs must therefore be able to be paused and resumed flexibly. Currently it is assumed, that switching between idle and normal operation happens instantaneously and can be performed as often as required. The number of these switching operation increases for lower utilisation values. In case of the \(\text{BAF}_\text{modern}\) setup's backfilling scenario the cluster had to pause and resume operation roughly 230 times in the simulated period with an average runtime interval length of \qty{20.7}{h} over the whole year. For practical application, however, a ramp-up and ramp-down time must definitely be taken into account. This might affect and shift the optimisation results. In addition, a more sophisticated scaling factor has to be implemented and replace the simple scaling via the number of logical cores, currently utilized in this study. Otherwise, a realistic consumption, especially for inhomogeneously structured computing clusters consisting of various CPU and accelerator modules is hard to estimate.


\section*{Acknowledgements}
The present analysis would not have been possible without the publicly available energy production and spot market price data by the Fraunhofer Institute for Solar Energy Systems (ISE). The authors would thus like to express their gratitude and support for open data.
In addition, the authors are grateful for access to the resources of the BAF2 cluster at the University of Bonn, the GridKa cluster at the Karlsruhe Institute of Technology, as well as the DEEP-System of the Jülich Supercomputing Centre to simulate the different setups.
This work was supported by the Federal Ministry of Research, Technology and Space (BMFTR) within the project FIDIUM (Förderkennzeichen 05H21PDRC1) and SUSFECIT (Förderkennzeichen 05D25PD4). The collaboration within the PUNCH4NFDI consortium in the NFDI (Deutsche Forschungsgemeinschaft (DFG, German Research Foundation) --- project number 460248186) was instrumental in carrying out the project. It was also supported by the environment of the cluster of excellence EXC~3107 \enquote{Color meets Flavor} (project number 533766364) as well as the collaborative research centre \enquote{NuMeriQS} (project number 511713970). We are very grateful for the support by all participants and collaborators of the SUSFECIT consortium, most importantly Markus Schumacher and Michael Böhler.


\printbibliography
\vspace*{3mm} 	



\end{document}